\documentstyle[12pt]{article}
\begin{document}
%\baselineskip=1.3\baselineskip
\baselineskip=1.2\baselineskip
\renewcommand{\thefootnote}{\fnsymbol{footnote}}
\setcounter{equation}{0}
\newcommand{\beq}{\begin{equation}}
\newcommand{\eeq}{\end{equation}}
\newcommand{\beqa}{\begin{eqnarray}}
\newcommand{\eeqa}{\end{eqnarray}}
\newcommand{\pr}{Phys. Rev.}
\newcommand{\pl} {Phys. Lett.}
\newcommand{\prl}{Phys. Rev. Lett.}

\pagestyle{plain}
\begin{titlepage}

{\flushright{  \vbox {IP/BBSR/96-34 \\ 
hep-th/9605163 }}}
\vspace{0.6cm}
\begin{center}
{\large{\bf {On the Fractional Electric Charge of a
Magnetic Monopole  at Nonzero Temperature}}}
\end{center}
\vskip 1.0cm
\begin{center}
{Alfred Goldhaber$^{a}$, \ Rajesh Parwani$^{b}$
\ and \ Harvendra Singh$^{b}$ \footnote{email :
goldhab@insti.phys.sunysb.edu; \
parwani@iopb.ernet.in ; \ hsingh@iopb.ernet.in}}
\end{center}
\vskip 0.6cm
\begin{center}
{$^a$ Institute for Theoretical Physics,\\
State Univerity of New York,\\
Stony Brook, NY 11794, \\ U.S.A. }
\end{center}
\vspace*{0.2cm}
\begin{center}
{$^b$ Institute of Physics \\ Sachivalaya Marg,
Bhubaneswar-751 005,  \\ INDIA}
\end{center}
\vskip 1.5cm
%\centerline{PACS 12.20.Ds, 52.60.+h, 11.15.Bt, 12.38.Mh.}
\centerline{{\it May 1996, { \bf Revised} June 1996}}
\vskip 1.5 cm
\centerline{\bf Abstract}

We extend
recent discussions about the effect of nonzero
temperature  on the induced electric charge, due to CP
violation, of a Dirac or an 't Hooft-Polyakov monopole.
In particular, we determine the fractional electric charge
of a very small 't Hooft-Polyakov monopole coupled to light
fermions at nonzero temperature.
If dyons with fractional electric charge  exist
in the Weinberg-Salam model, as recently suggested in the
literature, then their charge too should be temperature
dependent.
\end{titlepage}
%\section{}
Recently two papers \cite{CP,GS} analyzed the temperature ($T$)
dependence of the
fractional electric charge induced respectively on (i) a Dirac
monopole with CP violating boundary conditions on the fermions, and
(ii) an 't Hooft-Polyakov monopole (in the absence of fermions)
under the influence of a
CP-odd $\theta$-term in the Lagrangian. Some comments in
Ref.\cite{GS} appear to contradict those in Ref.\cite{CP}.
Our purpose
here is to resolve this apparent discrepancy and to extend
the study of Ref.[2] by including light fermions. We also note that
the result of Ref.[1] may be simply modified to accommodate a
chiral angle for the fermion masses.  \\

As shown by Schwinger and Zwanziger, Dirac's quantization
condition for monopoles \cite{Dirac} must be generalized in the
case of dyons to \cite{SZ}
\beq
e_1~g_2 - e_2 ~g_1 = {1\over2}\times \mbox{integer}
\label{1}
\eeq
where $( e_i, g_i),~1\le i \le2$, are the electric and magnetic
charges of the $i^{th}$ dyon. From the existence of electrons of
charge $(e,0)$, one deduces from (1) that the difference in the
electric charge $(q-q')$ of two dyons, carrying minimum magnetic
charge $g=1/2 e$, is an integral multiple of $e $,
\beq
q - q'= n\,e,~~~~~~~~~n\in Z.
\eeq
However in the absence of CP conservation, there is no further
restriction on $q$ or $q^{\prime}$ so that, in general, they may be
irrational.
Witten \cite{EW} considered CP violation due to a vacuum
angle term
\beq
\Delta{\cal L}= {\theta e^2\over 32 \pi^2} F^a_{\mu\nu} {\tilde
F}^{\mu\nu}_a
\eeq
in
the Lagrangian of a spontaneously broken gauge theory containing
't Hooft-Polyakov monopoles \cite{HP}. By canonical methods
it was shown
in \cite{EW} that at zero temperature the monopole acquires a
fractional electric charge
\beq
Q=- {e \theta\over 2 \pi} \; ,
\eeq
and that this result is exact.\\

In Ref.[2] the 't Hooft-Polyakov monopole was quantized by the
method of collective coordinates to leading order in
temperature $T$. Since fluctuations of fields were ignored in
[2], their calculation is at tree level in the fields. At that
order they obtained again Eq.(4). We believe the result may be
understood as follows. As is well known, the partition function
may be represented by an Euclidean path integral with
periodic boundary conditions in the time direction for
the bosons. The action is
\beq
S=\int_0^{1\over k T} d\tau \int d^3 x ~( {\cal L} + \Delta {\cal
L}) \; ,
\eeq
where ${\cal L}$ is the CP even part and $\Delta{\cal L}$ is
given by (3).
Therefore the CP-odd part of the static, effective
Lagrangian obtained from (5) is to leading order (when field
fluctuations are ignored), the same as (3) and hence the charge
must be given by the same expression (4) at nonzero temperature.
More physically, fluctuations may be ignored if
the  temperature is much below the scale for creation of free
charge carriers, or in this case the mass of a charged vector
boson.
Thus the conclusion  of Ref.[2] appears to be
essentially the result of an implicit zero temperature analysis.
\\

In order to study the change in the result of
Ref.[2] when light fermions\footnote{We shall
discuss only massive fermions here to avoid subtleties
involved with the massless case (see \cite{C,W,Y,YG} for a
discussion at zero temperature). More specifically we shall
restrict ourselves to the regime $M_W >> T$ and $M_W >> M$,
with $M_W$ the W-boson mass and $M$ the fermion mass. The
monopole size is $\sim 1/M_W$ and the temperature is well below the
symmetry restoration scale.} are coupled to the
monopole we first recall the situation at zero temperature :
Though the net charge is still given by Eq.(4)
when fermions are
coupled in a CP-invariant way to an 't Hooft-Polyakov monopole,
the distribution of charge changes significantly.
As noted by Callan and others \cite{C,W,Y} it is
energetically preferable for the charge to reside with the light
fermions rather than in the small monopole core.
Computationally one sees the problem as a breakdown of the
loop expansion in the limit of a
small monopole. In Ref.\cite{Y} a
variational calculation, which included the
Coulomb energy of the dyon, showed that in the limit of a point
monopole the energy of the system was minimized when
the { \it effective} boundary conditions for the fermions
violated CP.
The effective boundary conditions become labeled by the vacuum
angle
$\theta$ which appeared originally as a coupling constant
in the Lagrangian.
It is these effective CP violating boundary conditions which
allow the dyon charge to be carried by the light fermions.\\

The conclusion of Ref.\cite{Y} should hold also for
temperatures $T \ll \alpha  M_W$,
(where $\alpha$ is the gauge coupling) so that
the thermal properties of a very small 't Hooft
Polyakov monopole coupled to light fermions may be deduced from
those  of the  Dirac monopole studied in Ref.[1].
We therefore review
the main features of the Dirac problem:
It was discovered
long ago \cite{KYG,G} that unless particular boundary conditions
at the Dirac monopole
are imposed on the wavefunctions of the electron
Hamiltonian, the problem is ill defined.
Assuming a conventional fermion mass,
equivalent to that obtained by Yukawa coupling to a scalar field,
the most general boundary conditions are labeled by an angular
parameter
${\bar\theta}$ \cite{G} with CP invariance  holding only
for  ${\bar\theta} =0~{\rm or}~\pi$. In \cite{YG} it was found
that at zero
temperature, at one-loop, the Dirac monopole acquired an electric
charge
\beq
Q_D(T=0) = -{e {\bar\theta} \over 2\pi},~~~~~~~~~-\pi<{\bar\theta}< \pi
\label{dc}
\eeq
due to vacuum polarization. The generalization of (\ref{dc}) to
nonzero temperature was obtained, at one-loop, in Ref.[1],
\beq
Q_D(T) = -{e x\over \pi} \ \sin{\bar\theta} \ \sum_{n=0}^{\infty}{1\over
(2 n +1)^2 + x^2 +x \cos{\bar\theta} \ \sqrt{(2 n +1)^2 + x^2}} \; ,
\label{seven}
\eeq
where $x \equiv M/(\pi T)$. In particular at high temperature,
\beq
Q_D( T \gg M) = -{e M\over T} \ {\sin{\bar\theta}\over 8} \ + \
O\left({M\over T}\right)^2.
\label{eigth}
\eeq
Amusingly, the
angular nature of ${\bar\theta}$ is not manifest at zero
temperature  but becomes so at nonzero temperature.
Also at nonzero temperature, $Q_D$
vanishes smoothly for the CP even values
$\bar \theta=0~{\rm or}~\pi$.\\

Returning to the 'tHooft-Polyakov problem, if the fermion
is an isodoublet then, in the point limit of the monopole,
 two angles
\cite{Y} $\theta_1 = \theta \pm {\pi \over 2}$
and $\theta_2 = \pm
{ \pi \over 2}- \theta$ are required to label the
boundary conditions.
These angles correspond respectively to the two states of charge
$\pm e/2$ represented in the fermion and are the analog of the
angle $\bar\theta$ in the Dirac monopole-electron problem.
Combining the results of Refs.\cite{CP} and \cite{Y} we determine   the
charge $Q_{HP}(\theta,T)$, at nonzero temperature,
of the  point
't Hooft-Polyakov monopole-fermion  system to be
\beqa
Q_{HP}(\theta,T) &=& {1 \over 2} \left[ Q_D(\theta_1, T) -
Q_D(\theta_2, T) \right] \label{hpc1}, \\
&=& -{e x^2 \over 2\pi} \sin{2\theta} \ \sum_{n=0}^{\infty}
{  1  \over
[(2 n +1)^2 + x^2]^{3 \over 2} -
x^2 \sin^2{\theta} \ \sqrt{(2 n +1)^2 + x^2}} \nonumber \; , \\
&& \label{hpc2}
\eeqa
where $Q_D$ is given by Eq.(7).
Notice that the charge vanishes at $\theta = \pm \pi/2$ which
correspond to the CP invariant values of $\pm \pi$ for
$\theta_1$ and $\theta_2$ above. Also, because of the
cancellation in (\ref{hpc1}) between the two oppositely
charged states,
the leading term at high-temperature
is of order $\left({M \over T}\right)^2 \sin 2\theta$
compared to $\left({M \over T}\right) \sin\bar\theta$ for the
Dirac case in (8). At zero
temperature, (\ref{hpc2}) may be evaluated as described
in \cite{CP} to give
\beq
Q_{HP}(\theta,T=0) = -{e \theta  \over 2 \pi}
\eeq
in agreement with Ref.\cite{Y} and Witten's formula (4).\\

In summary, Witten's formula (4) changes at nonzero temperature.
An explicit computation has been possible in the limit of an
infinitely
massive monopole coupled to light fermions. The vacuum angle
$\theta$ of the theory
then becomes the parameter labeling the effective CP violating
boundary conditions at the location of the monopole. While the
result (9) above applies in the point limit, we expect most of
the
features discussed
to be qualitatively the same slightly away from this
limit.  However, for monopoles of sufficiently large
radius the above
analysis is sure to break down. Corrections may become
appreciable
already when the temperature $T$ is within an  order
of magnitude  of the vector boson mass  $M_W$.  Even though the
magnetic pole strength still is well defined at this point,
the appearance of charged boson pairs begins to be noticeable.\\

As in the Dirac case studied in \cite{CP}, we caution  that
at nonzero temperature the charge of the monopole is
a thermal expectation
value rather than an eigenvalue. Furthermore the charge,
which is localised within the fermion's thermal Compton wavelength, is Debye
screened at large distances. Clearly then it makes sense
to talk of the charge only if the fermion thermal Compton wavelength
is much smaller than the Debye length. This last condition is
satisfied, for example, at high temperature when $T >> eT >> M$ : so that
the Debye length $\sim 1/eT$ is much larger than the
fermion's thermal Compton wavelength $\sim 1/T$. \\

So far we have considered CP violation in the form
of a vacuum angle or through fermionic boundary conditions
(in the point limit above these were seen to be linked).
One can also consider  CP violation in the form of chiral angles
in the fermion masses. For example, if the electron mass $M$ is
replaced by $M e^{i \gamma_5 \omega}$ then,
as remarked in Ref.\cite{YG},
the  result (6) for the charge of a Dirac monopole
is modified by the replacement $\bar\theta \to
\bar\theta + \omega$. We have verified that this replacement
holds more generally for the nonzero temperature result
(7). \\

At zero temperature,
instead of the local or Gauss's law charge discussed up till now,
one can define also an Aharonov-Bohm (AB) or Lorentz-force
charge, as measured in an AB diffraction experiment. A
gauge invariance argument of Wilczek \cite{W}, elaborated by
Goldhaber, {\it et. al.} \cite{GMW}, showed that at zero
temperature the fractional AB charge of
a dyon is exactly equal to its vacuum angle
contribution.
In vacuum at zero temperature the AB and local charges
are identical.  Thus a
nonzero vacuum angle is a necessary as well as a sufficient
condition for
fractional dyon charge.
Since the vacuum angle may be removed
from the Lagrangian by a chiral rotation of the
fermions, this helps to justify the
resemblance of Eq.(6) to  Eq.(4).
At nonzero temperature,
as discussed above, the local charge is subject to
thermal fluctuations and leakage,
and also is screened at large distances. What happens to the AB
charge is less clear to us, but it looks quite likely that it
also is diminished and subject to
fluctuations because random collisions with
the quasiparticles in the plasma  would dephase
contributions from different spacetime paths.\\

Finally, we note that the Weinberg-Salam
model of electroweak interactions appears to have
some peculiar
magnetic monopole solutions \cite{N,V,CM}. So far their
properties have been studied in the absence of fermions. Since
fermions in this model are a source of CP violation,
it has been suggested \cite{V} that
these monopoles may become dyons carrying irrational
electric charges.
It would be fascinating if such dyons turn out to
be phenomenologically relevant. From our discussion here we
anticipate the electric charge of such an electroweak dyon to be
much smaller at high temperature than at zero
temperature.\\ \\

This work was supported in part by the U.S. National Science
Foundation.

\end{document}